\documentclass[aps,prl,twocolumn,groupedaddress]{revtex4-1}
\usepackage{graphicx}

\begin{document}


\title{Constraining annihilating dark matter by radio data of M33}


\author{Man Ho Chan}
\affiliation{The Education University of Hong Kong}


\date{\today}

\begin{abstract}
Recent studies of radio data put strong constraints on annihilation cross section for dark matter. In this article, we provide the first analysis of using M33 radio data in constraining annihilating dark matter. The resulting constraints of annihilation cross sections for some channels are more stringent than that obtained from 6 years of Fermi Large Area Telescope (Fermi-LAT) gamma-ray observations of the Milky Way dwarf spheroidal satellite galaxies. In particular, the conservative lower limits of dark matter mass annihilating via $e^+e^-$, $\mu^+\mu^-$and $\tau^+\tau^-$ channels are 190 GeV, 120 GeV and 70 GeV respectively with the thermal relic annihilation cross section. These results are in large tensions with some of the recent quantitative analyses of the AMS-02 and Fermi-LAT data of the Milky Way center.
\end{abstract}


\maketitle


\section{Introduction}
In the past few years, gamma-ray observations and positron detections revealed the possibility of dark matter annihilation in our galaxy. For example, the detection of positrons from the Alpha Magnetic Spectrometer (AMS) indicates that some excess positrons exist in the Milky Way, which might originate from non-standard astrophysical processes \cite{Aguilar,Accardo,Aguilar2}. Besides, some excess gamma-ray signal from the Milky Way center is observed by the Fermi Large Area Telescope (Fermi-LAT) \cite{Daylan,Abazajian}. The signal is observed to extend to at least $10^{\circ}$ from the Milky Way center, which is difficult to be explained by the emission of millisecond pulsars \cite{Daylan}. Therefore, some studies claim that the excess GeV gamma rays and the excess positrons detected can be best explained by dark matter annihilation \cite{Daylan,Abazajian,Calore,Calore2,Abazajian2,Boudaud}. Originally, detailed analyses based on the electron and positron data from AMS-02 suggest that the dark matter mass should be of the order $m \sim 100-1000$ GeV \cite{Boudaud}. For the Fermi-LAT data of the Milky Way center, the best-fit dark matter mass is $\sim 30-40$ GeV via the $b\bar{b}$ channel \cite{Daylan}. However, recent data from Fermi-LAT observations of the Milky Way dwarf spheroidal (MW dSph) satellite galaxies and some other indirect observations \cite{Ackermann,Geringer,Lopez,Chan} put very tight upper limits on the annihilation cross section. Most of the limits are lower than the best-fit ranges of the dark matter interpretation of the gamma-ray and positron excess. Therefore, the above suggested proposals are now disfavored. 

Later, some new analyses suggest that $m \sim 10-100$ GeV dark matter annihilating via $e^+e^-$, $\mu^+\mu^-$ or $\tau^+\tau^-$ channel can also obtain good fits to the AMS-02 data \cite{Mauro}. Also, by relaxing the dark matter density profile for the Milky Way, \cite{Abazajian2} obtain new sets of parameters that can explain the gamma-ray excess. These claims can marginally satisfy the Fermi-LAT MW dSph constraints. However, some other studies based on radio data seem to disfavor these claims \cite{Egorov,Cholis,Chan2}. In this article, we use the radio data of M33 to test the claims and obtain some new constraints for annihilating dark matter via six standard model annihilation channels. This is the first time to use the radio data of M33 to constrain annihilating dark matter. We will use the latest data and parameters to calculate the new constraints. 

\section{Radio observations of M33}
The group in \cite{Tabatabaei} uses radio telescope to observe 3.6 cm and 20 cm emission of M33 spiral galaxy. After analyzing the radio data, a significant amount of non-thermal emission originated from a large region $R \le 7.5$ kpc is detected. The non-thermal radio flux density for 3.6 cm ($\nu=8.3$ GHz) and 20 cm ($\nu=1.5$ GHz) observations are $370 \pm 137$ mJy and $2245 \pm 145$ mJy respectively \cite{Tabatabaei}. In fact, the non-thermal emission can originate from star formation in M33 or some other astrophysical processes. Nevertheless, if we assume that all the non-thermal radio radiation originates from synchrotron radiation of the electron and positron pairs produced by dark matter annihilation, the upper limits of the above two fluxes can be used to give conservative constraints of dark matter annihilation. The upper bounds of the total energy flux are $S \le 4.2 \times 10^{-14}$ erg cm$^{-2}$ s$^{-1}$ (for 3.6 cm) and $S \le 3.6 \times 10^{-14}$ erg cm$^{-2}$ s$^{-1}$ (for 20 cm).

The magnetic field strength of M33 is $B=8.1 \pm 0.5$ $\mu$G \cite{Berkhuijsen}. This large magnetic field strength ensures that the synchrotron radiation dominates the cooling rate of the electron and positron pairs. We also include another major cooling process, the inverse Compton scattering, in our calculations. It contributes 15\% of the total cooling rate, and it would suppress the resulting radio signal slightly. The diffusion time scale of the electron and positron pairs is much longer than the cooling time scale. If we take a conservative value of the diffusion coefficient $D_0 \sim 10^{28}$ cm$^2$ s$^{-1}$ for M33 \cite{Berkhuijsen}, the diffusion and cooling time scales are $t_D \sim R^2/D_0 \sim 5 \times 10^{16}$ s and $t_c \sim 1/b \sim 5 \times 10^{15}$ s respectively, where $b \approx 2 \times 10^{-16}$ GeV s$^{-1}$ is the total cooling rate \cite{Colafrancesco}. Based on this result, the diffusion term in the diffusion equation can be neglected and the injected spectrum of the electron and positron pairs is proportional to the source spectrum of dark matter annihilation $dN_e/dE$ \cite{Storm}.

By using the monochromatic approximation \cite{Bertone}, the total synchrotron radiation energy flux of the electron and positron pairs produced by dark matter annihilation is given by \cite{Bertone,Profumo}:
\begin{equation}
S \approx \frac{1}{4 \pi d^2} \left[ \frac{6.78<\sigma v>}{m^2} \int_0^R 4 \pi r^2 \rho^2EY(m,E)dr \right],
\end{equation}
where $d=840$ kpc \cite{Seigar}, $\rho$ is the dark matter density profile of M33, $<\sigma v>$ is the annihilation cross section, $E=0.43(\nu/{\rm GHz})^{1/2}(B/{\rm mG})^{-1/2}$ GeV, and $Y(m,E)=\int_E^m(dN_e/dE')dE'$. The electron or positron spectrum $dN_e/dE$ for each annihilation channel can be obtained in \cite{Cirelli}. Here, we assume that the magnetic field strength in M33 is uniform for $r \le 7.5$ kpc. In fact, the resulting energy flux does not sensitively depend on $B$. Also, we expect that the magnetic field near the center of M33 is much higher than 8.1 $\mu$G. Therefore, using the constant profile of $B$ would give an underestimated total radio flux $S$. In other words, this can give conservative constraints for dark matter annihilation cross section.  

In the following, we use two profiles, the Navarro-Frenk-White (NFW) profile \cite{Navarro} and the Burkert profile \cite{Burkert}, to model the dark matter density profile of M33. The latest parameters of these profiles based on dynamical analysis of the M33 rotation curve are $(\rho_s,r_s)=(1.79^{+0.77}_{-0.61},6.42^{+0.56}_{-0.47})$ and $(\rho_c,r_c)=(12.1 \pm 2.0,7.5 \pm 1.5)$ \cite{Fune,Corbelli} respectively, where the units for the densities and scale radii are in $10^{-25}$ g cm$^{-3}$ and kpc. By putting different density profiles into Eq.~(1) with the above parameters, we can get an analytic expression for the lower limit of $S$:
\begin{equation}
S \ge S_0 \left( \frac{<\sigma v>}{2.2 \times 10^{-26}~\rm cm^3~s^{-1}} \right) \left(\frac{m}{\rm GeV}\right)^{-2}Y(m,E),
\end{equation}
where $S_0$ is a constant which depends on the density profile and the characteristic peak energy $E$ (see table 1).

\begin{table}
\caption{The values of $S_0$ (in erg cm$^{-2}$ s$^{-1}$) for different dark matter density profiles and radio observations.}
 \label{table1}
 \begin{tabular}{@{}lcc}
  \hline
   & 3.6 cm ($E=13.8$ GeV) & 20 cm ($E=5.85$ GeV) \\
  \hline
  $S_0$ (NFW) & $2.6^{+1.6}_{-1.1} \times 10^{-10}$ & $11.2^{+6.6}_{-5.0} \times 10^{-11}$  \\
  $S_0$ (Burkert) & $1.4^{+0.1}_{-0.1} \times 10^{-10}$ & $6.0^{+0.4}_{-0.3} \times 10^{-11}$ \\
  \hline
 \end{tabular}
\end{table}

From table 1, we can see that both density profiles give similar radio flux. Therefore, the choice of the density profiles is not a key factor in this analysis. We will use the NFW profile to do the analysis as it gives better fits to the rotation curve data of M33 \cite{Corbelli}. By using the data of $dN_e/dE$ in \cite{Cirelli}, we can calculate $Y(m,E)$ for different $m$ and the two characteristic energies ($E=5.85$ GeV for the 20 cm data and $E=13.8$ GeV for the 3.6 cm data). By comparing Eq.~(2) with the observed upper bounds of the total non-thermal radio flux, we can get the upper limit of the annihilation cross section for each channel (see Fig.~1). The resulting constraints of annihilation cross sections for two annihilation channels ($e^+e^-$ and $\mu^+\mu^-$) are generally more stringent than that obtained from 6 years of Fermi-LAT gamma-ray observations of the MW dSph satellite galaxies \cite{Ackermann}, while the other four channels give less stringent constraints. It is not a surprising result because the electrons and positrons produced by these four channels are generally dominant in the lower energy range (e.g. most positrons are at $E \le 5$ GeV for $m=50$ GeV via $b\bar{b}$ channel). However, the radio data used are focusing on higher energies ($E=5.85$ GeV and $E=13.8$ GeV) positions. Therefore, based on this limitation, we will mainly focus on the $e^+e^-$ and the $\mu^+\mu^-$ channels and use these results to compare with the existing models. We will also use the results of the $\tau^+\tau^-$ and $b\bar{b}$ channels as we will see soon they can also give tight constraints if we consider the effect of substructures. 

\begin{figure}
\vskip 10mm
 \includegraphics[width=82mm]{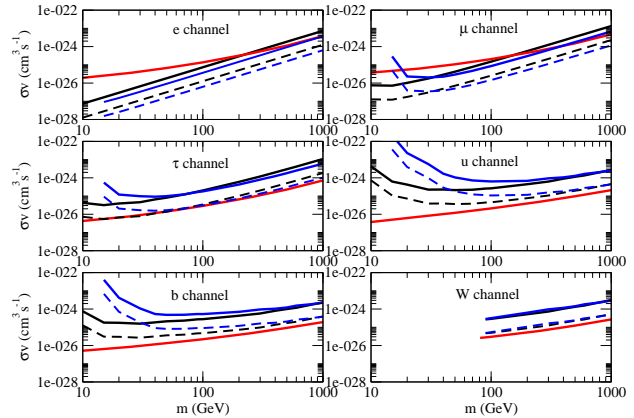}
 \caption{The upper limits of the annihilation cross sections for the six annihilation channels. The black and blue solid lines represent the upper limits obtained by 20 cm and 3.6 cm radio data respectively (without substructure contribution). The dashed lines represent the corresponding upper limits with the consideration of substructures. The red solid lines represent the upper limits obtained by Fermi-LAT gamma-ray observations of Milky Way dwarf spheroidal satellite galaxies \cite{Ackermann}.}
\vskip 10mm
\end{figure}

In standard cosmology, the thermal relic annihilation cross section should be $<\sigma v>=2.2 \times 10^{-26}$ cm$^3$ s$^{-1}$ for $m \ge 10$ GeV if dark matter particles are thermal relic particles \cite{Steigman}. This value is determined by the total dark matter content in our universe. We can see that for the channels $e^+e^-$ and $\mu^+\mu^-$, the allowed ranges of $m$ are $m \ge 80$ GeV and $m \ge 40$ GeV respectively for the thermal relic annihilation cross section. However, for the other four channels, the upper bounds of the cross sections are generally above the thermal relic annihilation cross section for $m \ge 10$ GeV.

\section{The effect of substructure}
Recent studies indicate that substructures would be formed when dark matter particles form structures. These substructures can greatly increase the annihilation rate to about an order of magnitude \cite{Strigari,Beck}. The boost factor $B_f$ can describe how substructures affect the total annihilation rate \cite{Strigari,Moline}. Recent simulations indicate that the boost factor can be simply obtained by the following empirical formula \cite{Moline}:
\begin{equation}
\log B_f= \sum_{i=0}^5b_i \left( \log \frac{M}{M_{\odot}} \right)^i,
\end{equation}
where $M$ is the virial mass of the structure and $b_i$ is the fitted coefficients \cite{Moline}. By using the virial mass of M33 $M=(4.3\pm 1.0)\times 10^{11}M_{\odot}$ \cite{Fune} and the most conservative model in the simulations \cite{Moline}, the boost factor for M33 is $B_f=4.86$. This value is reasonable as the size of M33 is in between a typical dwarf galaxy and a typical normal sized galaxy (e.g. $B_f=3.4$ for Draco dwarf galaxy, $B_f=10$ for M81 galaxy, $B_f=35$ for Coma cluster \cite{Beck}). If we take this boost factor and re-calculate the upper limits of the annihilation cross sections, we can get tighter constraints (see Fig.~1). The allowed ranges for the channels $e^+e^-$, $\mu^+\mu^-$ and $\tau^+\tau^-$ are $m \ge 190$ GeV, $m \ge 120$ GeV and $m \ge 70$ GeV respectively for the thermal relic annihilation cross section. These results are similar to our previous results using the radio data of M31 \cite{Chan2}.

\section{Comparing the results with the latest fits of the positron and gamma-ray data}
Recent re-analysis of the AMS-02 data further decreases the best-fit mass and cross sections to $\sim 10-100$ GeV and $<\sigma v> \sim 10^{-26}-10^{-24}$ cm$^3$ s$^{-1}$ \cite{Mauro}. Since the uncertainties of the fits for the $b\bar{b}$ and $W^+W^-$ channels are very large, we just focus on the other three channels $e^+e^-$, $\mu^+\mu^-$ and $\tau^+\tau^-$. The best-fit parameters for these three channels are given in table 2.

\begin{table}
\caption{The best-fit parameters of the AMS-02 data for three channels \cite{Mauro}.}
 \label{table2}
 \begin{tabular}{@{}lcc}
  \hline
   Channel & $m$ & $<\sigma v>$ \\
  \hline
  $e^+e^-$ & $50^{+1}_{-4}$ GeV & $5.6^{+2.2}_{-2.6} \times 10^{-27}$ cm$^3$ s$^{-1}$ \\
  $\mu^+\mu^-$ & $88^{+31}_{-9}$ GeV & $7.9^{+12.6}_{-3.4} \times 10^{-26}$ cm$^3$ s$^{-1}$ \\
  $\tau^+\tau^-$ & $635^{+73}_{-195}$ GeV & $7.2^{+1.4}_{-3.5} \times 10^{-24}$ cm$^3$ s$^{-1}$ \\
  \hline
 \end{tabular}
\end{table}

Without considering the substructure contribution, most of the best-fit parameters are allowed except for the $\tau^+\tau^-$ channel. However, if we consider the substructure effect, all these parameters are in large tensions with our results (see Fig.~2). In fact, our previous study using the radio data of M31 tends to disfavor these parameters \cite{Chan2}. Now, our results based on the radio data of M33 also support this claim. The dark matter interpretation (with the annihilation channels $e^+e^-$, $\mu^+\mu^-$ and $\tau^+\tau^-$) of the positron excess based on the AMS-02 data should be ruled out.

Besides the data of positron excess, \cite{Abazajian2} also analyze the Fermi-LAT data of the Milky Way center and obtain some best-fit parameters to account for the gamma-ray excess in the Milky Way. By adopting the best models for the Milky Way halo and background diffuse emission models, they obtain $m=43^{+2.1}_{-1.9}\pm 19$ GeV and $<\sigma v>=7.4^{+2.7}_{-2.3} \times 10^{-26}$ cm$^3$ s$^{-1}$ for the $b\bar{b}$ channel and $m=9^{+0.27}_{-0.23} \pm 2$ GeV and $<\sigma v>=2.2^{+1.2}_{-0.7} \times 10^{-26}$ cm$^3$ s$^{-1}$ for the $\tau^+\tau^-$ channel \cite{Abazajian2}. In Fig.~2, we can see that these parameters are in large tensions with our results too. Therefore, the dark matter interpretation of the gamma-ray excess in the Milky Way is also dubious. 

\begin{figure}
\vskip 10mm
 \includegraphics[width=82mm]{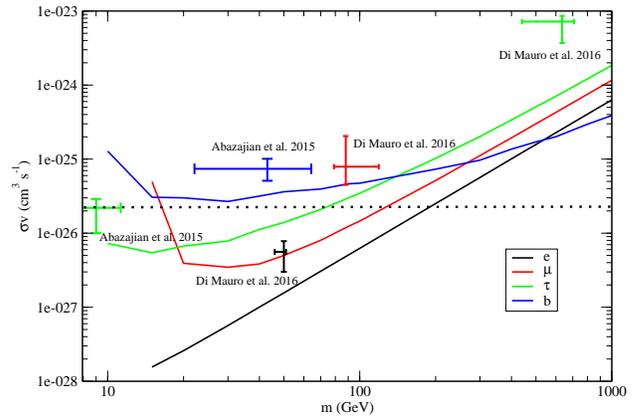}
 \caption{The upper limits of the annihilation cross sections for the $e^+e^-$, $\mu^+\mu^-$, $\tau^+\tau^-$ and $b\bar{b}$ channels. The data points with 1$\sigma$ error bars are the results obtained in \cite{Abazajian2,Mauro} for the dark matter interpretation of the GeV excess and positron excess respectively. The dotted line is the thermal relic annihilation cross section.}
\vskip 10mm
\end{figure}

\section{Discussion}
The M33 galaxy is a good candidate for constraining annihilating dark matter. It is because M33 is a nearby galaxy and most of its parameters are well-determined. Also, it is believed that the supermassive black hole at M33 is not very massive ($M_{BH} \le 1500M_{\odot}$ \cite{Seigar}) so that the resulting radio signal would not be affected significantly due to this source. In this article, we provide the first analysis of using M33 radio data to derive the radio constraints of annihilating dark matter. Without considering the substructure effect, our results give more stringent constraints on annihilation cross sections for two annihilation channels, $e^+e^-$ and $\mu^+\mu^-$. The allowed ranges of dark matter mass for these two channels are $m \ge 80$ GeV and $m \ge 40$ GeV respectively for the thermal relic annihilation cross section. If the effect of substructures is considered, the allowed ranges would be tightened to $m \ge 190$ GeV and $m \ge 120$ GeV respectively. Also, for the $\tau^+\tau^-$ channel, the viable range is $m \ge 70$ GeV. 

We compare our results with the best-fit parameters obtained from recent analyses of the AMS-02 data and the Fermi-LAT data of the Milky Way center. We find that our results generate large tensions with all these best-fit parameters (except for the $b\bar{b}$ and $W^+W^-$ channels in the analysis of the positron excess). Our results generally agree with our previous analysis by using the radio data of M31 \cite{Chan2}. Therefore, the dark matter interpretation of the positron excess and the gamma-ray excess is highly dubious. In fact, many studies point out that the positron excess and gamma-ray excess may originate from some astrophysical processes \cite{Delahaye,Mauro2,Yuan,Petrovic,Gaggero,Brandt,Bartels}. Further observations based on new instruments (e.g. DAMPE mission \cite{Gargano}) may be able to verify our claims.

\section{acknowledgements}
This work is supported by a grant from The Education University of Hong Kong (Project No.:RG4/2016-2017R).

\end{document}